\documentclass[twocolumn,floats,prl,aps]{revtex4}

\usepackage{graphicx}

\newcommand\beq{\begin{equation}}
\newcommand\eeq{\end{equation}}
\newcommand\bea{\begin{eqnarray}}
\newcommand\eea{\end{eqnarray}}
\newcommand{\nonum}{\nonumber}

\begin{document}

\title{\bf Non-universal tunneling resistance at the quantum critical point of   
mesoscopic SQUIDs array}

\author{\bf Sujit Sarkar}
\address{\it
PoornaPrajna Institute of Scientific Research,
4 Sadashivanagar, Bangalore 5600 80, India.\\
}
\date{\today}


\date{\today}

\begin{abstract}
We calculate the
tunnelling resistance at the quantum critical point
of a mesoscopic
SQUIDs array
in the presence of magnetic flux. 
We find the analytical relation between the magnetic flux induced 
dissipation strength and
the Luttinger liquid parameter of the system. While the experimental finding
for the system is around $40-50$ mK, we find the behavior of the system 
even at lower temperatures through the analysis of renormalization group. 
Apart from the length scale dependent superconductor-insulator transition,
we also predict the evidence of length scale independent metallic state. 
This study also emphasizes the importance
of Co-tunelling effect.  
\vskip .4 true cm
\end{abstract}
\maketitle


\noindent
{\bf Introduction:} 
Josephson junction arrays have attracted considerable interest in the recent
years owing to their interesting physical properties. Currently such arrays can
be fabricated in restricted geometries both in one and two 
dimensions 
\cite{sondhi,suj1,suj2,granto1,geer,zant,kuo,havi2,havi3,lar}.
There are a few experimental findings in the field of mesoscopic SQUIDs array.
The authors of Ref. (\cite{havi3} ) and Ref. (\cite{kuo}) 
have fabricated the arrays of SQUIDs 
junctions of different numbers in a single chip. 
They have studied the current voltage
characteristics of the mesoscopic SQUIDs array in the presence of a magnetic field.
They have used an external magnetic field $(B)$ that tunes the
effective Josephson coupling ($ {E_J}$) 
between the nearest neighbor superconductors
by the following relations 
$ {E_J }= {E_ {J0} } |cos (\frac{\pi \Phi}{{\Phi}_{0}})| $,
where $\Phi $ is the external flux and ${\Phi}_0 $ is the flux 
quantum. 
They have found magnetic field induced superconductor-insulator (SI) quantum phase
transition
at around $40-50$ mK. When the applied magnetic flux is less than the
critical value the system shows constant saturated resistance that may arise from
the source of quantum phase slip centers (QPS) \cite{suj2}. 
When the magnetic flux exceeds the
critical value the system turns into the insulating phase due to the flux induced
Coulomb blockade of Cooper pair tunneling. 
In this letter, we study the behavior of the system at very low 
temperature where there is
no experimental findings, i.e., around few
mili Kelvin \cite{havi3,kuo}. 
The experimental result shows length scale dependent 
superconductor-insulator (SI)
transition at the quantum resistance but this is not the 
whole picture, we find the length scale independent metallic phase.  
We calculate the tunneling resistance at the quantum critical point 
and  also show explicitly the importance of Co-tunneling effect 
to get the correct physical behavior of the system.\\  
{\bf Analytical relation between the flux induced 
dissipative strength and the Luttinger liquid
parameter:} 
Here we derive analytical expression of magnetic flux induced 
dissipative strength ($\alpha $) in
terms of the interactions of the system. 
At first we derive the dissipative action/partition function 
of a
quantum impurity system. We will see that the analytical structure of this
dissipative action is identical with the mesoscopic SQUIDs array. 
\\
Here we consider that the impurity is present at the origin where the fermions
scatter from the left to the right and vice versa. The Hamiltonian describing
this process is
$ {H_1} =  {V_0} ( {R^{\dagger}} (0) {L} (0) + h.c ) 
= {V_0} \int dx {\delta} (x) cos{\theta} (x) . $ 
The total Hamiltonian of the system 
$ H = {H_0}~+~ {V_0} \int dx  {\delta} (x) cos{\theta} (x,\tau)$
$ {H_0} = \frac{1}{2 \pi} \int { u K {({{\partial}_x} {\theta}(x,\tau) )}^{2} 
~+~ \frac{u}{K}
{({{\partial}_x} {\phi}(x,\tau ) )}^{2} }, $
corresponding Lagrangian of the system is
\bea
{L} & = & \frac{1}{2 \pi K} \int { \frac{1}{ u} 
{({{\partial}_{\tau}} {\phi}(x,\tau) )}^{2} ~+~ {u}
{({\partial} {\phi}(x,\tau) )}^{2} } \nonum\\ 
& & + {V_0} \int dx  \delta (x) cos(\theta (x,\tau)) 
= {L_0} + {L_1} 
\eea
where ${L_0}$ and ${L_1}$ are the non-interacting and the interacting part
of the Lagrangian and $K$ is the Luttinger liquid parameter of the system.
The only non-linear term in this Lagrangian is expressed by the field 
$\theta (x=0 )$. We would like to express the action of the system as an
effective action by integrating the field $\theta (x \neq 0 )$. Therefore
one may consider $\theta (x \neq 0 )$ as a heat bath, which yields the
source of dissipation in the system. The constraint condition 
for the integration is
$ {\theta} ( \tau ) = {\theta} (x=0, \tau )$. We can write the partition function.
\bea
Z & = & \int D \theta (x,\tau) e^{- \int_{0}^{\beta} L d {\tau}} \nonum\\
& & = \int  D \theta (x,\tau)  D \theta (\tau) 
\delta (\theta (\tau) - {\theta} (0, \tau ) ) e^{- \int_{0}^{\beta} L d {\tau}}  
\eea
Here we use the standard trick of introducing the Lagrange multiplier with
auxiliary field $\lambda (\tau)$.
$ Z  =  \int {D \theta (x,\tau)} \int {D \theta (\tau)} \int D {\lambda (\tau)} 
e^{- \int_{0}^{\beta} ({ L_0 + L_1 }) d {\tau}} $\\ 
$  e^{i \int_{0}^{\beta} d {\tau} {\lambda (\tau)} 
~(\theta (0, \tau) - {\theta} (\tau ) )} $\\
$ Z  =  \int {D \theta (\tau)} e^{- \int_{0}^{\beta}  {L_1}  d {\tau} }
\int D {\lambda (\tau)} e^{-i {\lambda (\tau)} {\theta} (\tau )} $ \\  
$ \int {D \theta (x, \tau)} e^{ \int_{0}^{\beta} (- { L_0} 
+ i {\lambda} ({\tau}) {\theta} (0, \tau) ) {d \tau} } $ \\
The Fourier transform of the first term of Eq.(2) is
$ {L_0} ~=~  \sum_{q} \sum_{i {\omega}_n } 
\frac{ {{\omega}_n}^{2}~+~ {v}^{2} {q}^{2} }{2 \pi K v} 
{\theta} (q, i {{\omega}_n} ) {\theta} (- q, - i {{\omega}_n} ) $ \\
At first we would like to calculate the integral: 
$ \int_{0}^{\beta} d {\tau} [{L_0}
- i {\lambda (\tau)} {\theta (0, \tau)} ] $, 
we can write this term as
$
\sum_{q} \sum_{i {\omega}_n }
\frac{ {{\omega}_n}^{2}~+~ {v}^{2} {q}^{2} }{2 \pi K v}
{\theta} (q, i {{\omega}_n} ) {\theta} (- q, - i {{\omega}_n} )
$
$  -\frac{1}{2 \sqrt{L} } ( {\lambda} (i {\omega}_{n}) 
 {\theta} (- q, - i {{\omega}_n}) + {\lambda} (- i {\omega}_{n})
 {\theta} (q, i {{\omega}_n}) $.
This integral appears in the integral $ \theta (x, \tau)$. This integral
is quadratic in $\theta $. Now we would like to perform the Gaussian 
integration by completing the square. We can write the result as
$ \frac{-1}{2 L} \sum_{i {{\omega}_n}, q} \frac{\pi K v}
{ {{\omega}_n}^{2}~+~ {v}^{2} {q}^{2} } $.\\
In the infinite length limit one can write,
$ \frac{1}{2 L} \sum_{q} \frac{\pi K v}
{ {{\omega}_n}^{2}~+~ {v}^{2} {q}^{2} }~=~ 
\int \frac{dq}{2 \pi} \frac{\pi K v}
{ {{\omega}_n}^{2}~+~ {v}^{2} {q}^{2} } = \frac{\pi K}{4 {{\omega}_n} } $.
Now we would like to append this result of integration in the second integral
of $Z$, i.e., the integral over ${\lambda}$. One can write the integrand 
as 
$\sum_{i {{\omega}_n } } ( - \frac{\pi K}{4 {{\omega}_n} } 
{\lambda} (i {\omega}_{n}) {\lambda} (-i {\omega}_{n})$ 
$ + \frac{i}{2} ( {\lambda} (i {\omega}_{n})
 {\theta} (- q, - i {{\omega}_n}) + {\lambda} (- i {\omega}_{n})
 {\theta} (q, i {{\omega}_n}) .$
This integral is again the quadratic integral of $\lambda $, therefore the
Gaussian integral can be performed by completing the square. We get after
the integration 
$ \sum_{i {{\omega}_n } } \frac{{\omega}_n}{\pi K} 
\theta (i {\omega}_{n}) \theta (- i {\omega}_{n}) $.
From these analytical expression, 
we obtain the effect of bath on ${\theta} (\tau) $. The appearance
of the factor $ {{\omega}_n} $ signifies the dissipation. Therefore the effective
action reduces to
\beq
S ~=~ \sum_{i {{\omega}_n } } \frac{{\omega}_n}{\pi K}
\theta (i {\omega}_{n}) \theta (- i {\omega}_{n}) +  
\int dx {V_0}  cos{\theta} (\tau)  
\eeq 
The above action implies that a single particle moving in the potential
$ {V_0} cos{\theta} (\tau)$ subject to dissipation with friction constant
, $\frac{1}{\pi K}$.  

Now we calculate the dissipative action of mesoscopic SQUIDs
array. We have already proved in Ref. \cite{suj2} that the strong coupling phase of
the system is consistent with the experimental findings. Here we calculate 
the effective partition function of our system in the strong coupling phase. Our
starting point is the Calderia-Legget \cite{cal} formalism.
Following reference we write the action as 
\beq
S_1 ~=~ S_0 ~+~ \frac{{\alpha}'}{4 \pi T}~ \sum_{m} {{\omega}_m}
{|{\theta}_m|}^2 .
\eeq
Here, $S_1$ is the standard action for the system with tiled wash-board
potential \cite{sch1,su1,schon}
to describe the dissipative physics for low dimensional superconducting
tunnel junctions,
$S_0 $ is the action for non-dissipative part,
${\alpha}' ~=~ \frac{R_Q}{R_s} cos|\frac{\pi \phi}{{\phi}_0}|$ (the
extra cosine factor which we consider in ${\alpha}'$ is entirely new in the
literature to probe the effect of an external magnetic flux and
is also consistent physically),
the Matsubara frequency
${\omega}_m ~=~ \frac{2 \pi}{\beta} m$ and $R_Q$
($ = 6.45 k \Omega$) is the quantum resistance and $R_s$ is
the tunnel junction resistance, $\beta$ is the inverse temperature.
In the strong potential, tunneling between the minima
of the potential is very small.
In the imaginary time path integral formalism,
tunneling effect in the strong coupling limit can be described
in terms of instanton physics.
In this
formalism, 
it is convenient
to characterize the profile of $\theta$ in terms of its time derivative,
\beq
\frac{d \theta {( \tau)} }{d {\tau} }~=~\sum_{i} e_i h (\tau - {\tau}_i),
\eeq
where  $h (\tau - {\tau}_i)$ is the time derivative at time $\tau$ of
one instanton configuration.
${\tau}_i$ is the location of the i-th instanton, $e_i = 1$ and $-1$
is the topological charge of instanton and anti-instanton respectively.
Integrating the function over $h$ from
$-\infty$ to $\infty$,
$ \int_{-\infty}^{\infty} d \tau h (\tau) = {\theta} (\infty)
- {\theta}({-\infty}) = 2 \pi . $
It is well known that the instanton (anti-instanton)
is almost universally constant except for a very small region of
time variation.
In the QPS process the amplitude of the superconducting
order parameter is zero only in a very small region of space as a function
of time and the phase changes by $\pm 2\pi$. 
So our system reduce to a neutral system consisting of equal number of
instanton and anti-instanton.
One can find the expression
for ${\theta} ( \omega)$, after the Fourier transform to the both sides
of Eq. 5 which yields
$ {\theta} ( \omega ) = \frac{i}{\omega} \sum_{i} e_i h (i \omega)
e^{i {\omega} {\tau}_1 } $
. Now we substitute this expression for ${\theta} ( \omega )$ in the
second term of Eq.4 and finally we get this term as
$ \sum_{ij} ~F({\tau}_i  - {\tau}_j ) {e_i} {e_j} $, where
$ F ({\tau}_i  - {\tau}_j ) = \frac{\pi \alpha}{\beta }
\sum_{m }~ \frac{1}{{ |{\omega}_m}|} e^{i {\omega} ({\tau}_i  - {\tau}_j )} $
$ \simeq  ln ({\tau}_i - {\tau}_j )$.
We obtain this expression for very small values of
${\omega}$ ( $\rightarrow 0 $).
So $ F ({\tau}_i  - {\tau}_j )$ effectively represents the Coulomb
interaction between the instanton and anti-instanton.
This term is the main source of dissipation physics
of the system. Following
the standard prescription of imaginary time path integral formalism, we can
write the partition function of the system as
\cite{suj2,lar,gia1,kane,zai1,furu}.
\bea
Z & = & \sum_{n=0}^{\infty} \frac{1}{n !} 
{z}^{n} \sum_{e_i} \int_{0}^{\beta} d {{\tau}_n}
\int_{0}^{{\tau}_{n-1}}d {\tau}_{n-1} ... \nonum\\
& & \int_{0}^{{\tau}_2} d {{\tau}_1}
e^{-F ({\tau}_i  - {\tau}_j ) {e_i}{e_j} }.
\eea
We would like to express the partion function in terms of integration over
auxiliary field, $ q{( \tau )}$. After some extensive analytical calculations, 
we get 
\beq
Z = \int D q ( \tau ) e^{(- \sum_{i {\omega}_n }
 \frac{ |{ {\omega}_n}|}{4 \pi \alpha} q (i {\omega}_{n}) q (- i {\omega}_{n}) )
 {+ (2 z \int_{0}^{\beta} d {\tau} cosq ( \tau ) )} } .
\eeq
Thus by comparing the first term of the action of Eq. (3) and the first term of 
exponential of Eq. (17), we conclude 
that the dissipative strength $\alpha$ and the Luttinger liquid parameter
of the system are related by the relation, $K = 4 \alpha $. 
\\
{\bf Quantum field theoretical study of model Hamiltonian of the system
and explicit derivation of dissipative strength:}\\
In our previous study, we have shown explicitly that the mesoscopic SQUIDs array
is equivalent to the array of superconducting quantum dots (SQD) with modulated
Josephson coupling.  
We first write the model Hamiltonian of SQD with nearest neighbor (NN)
Josephson coupling and
also with the presence of the
on-site and NN charging energy between SQD,
\beq
H~=~H_{J1}~+~H_{EC0}~+~H_{EC1}.
\eeq
Now we would like to recast our basic Hamiltonians in the spin
language. This is valid when ${E_{C0}} >> {E_{J1}}$.
It is also observed from the
experiments that the quantum critical point exists for 
larger values of the magnetic field, when the magnetic field
induced Coulomb blockade phase is more prominent than the
$E_J $ induced SC phase.
Thus our theoretical model
is consistent with the experimental findings.
During this mapping process we follow Ref. (\cite{lar,suj1,suj2}).
$
H_{J1}~=~ -2~E_{J1} \sum_{i}
( {S_i}^{\dagger} {S_{i+1}}^{-} + h.c)
$,
$
H_{EC0}~=~ {E_{C0}} \sum_{i}
{S_i}^{Z}.$
$
H_{EC1}~=~4 E_{Z1} \sum_{i} {S_i}^{Z}~{S_{i+1}}^{Z},
$
$ {E_{J1} }= {E_ {J10} } |cos (\frac{\pi \Phi}{{\Phi}_{0}})| $
. At the Coulomb blocked regime, the higher order expansion leads
to the virtual state with energies exceeding $E_{C0}$. 
In this second order process, the effective Hamiltonian reduces to
the subspace of charges $0$ and $2$, and takes the form
\cite{lar,suj1,suj2},
\beq
H_C ~=~- \frac{3 {E_{J1}}^2 }{4 E_{C0}} \sum_{i}
{{S_i}^Z}{{S_{i+1}}^Z} ~-~ \frac{{E_{J1}}^2}{E_{C0}}
\sum_{i} ({S_{i+2}}^{\dagger} {S_i}^{-} + h.c).
\eeq
With this corrections $H_{C1}$
become
$
H_{EC1}~\simeq~ (4 E_{Z1}~-~\frac{3 {E_{J1}}^2}{4 E_{C0}})
~\sum_i {{S_i}^Z}{{S_{i+1}}^Z}
$
One can express
spin chain systems to as spinless fermions systems through
the application of Jordan-Wigner transformation. We have
transformed all Hamiltonians in spinless fermions which we
have not present in this letter. 
In order to study the continuum field theory of these Hamiltonians,
we recast the spinless
fermions operators in terms of field operators by a relation \cite{gia1}.
$ {\psi}(x)~=~~[e^{i k_F x} ~ {\psi}_{R}(x)~+~e^{-i k_F x} ~ {\psi}_{L}(x)] $
where ${\psi}_{R} (x)$ and ${\psi}_{L}(x) $ describe the second-quantized
fields of right- and
the left-moving fermions respectively.
We would like to express the fermionic fields in terms of bosonic
field by the relation
$ {{\psi}_{r}} (x)~=~~
\frac{U_r}{\sqrt{2 \pi \alpha}}~~e^{-i ~(r \phi (x)~-~ \theta (x))} $,
$r$ is denoting the chirality of the fermionic fields,
right (1) or left movers (-1).
The operators $U_r$ preserve the anti-commutivity of fermionic fields. 
$\phi$ field corresponds to the
quantum fluctuations (bosonic) of spin and $\theta$ is the dual field of $\phi$. 
They are
related by the relations
$ {\phi}_{R}~=~~ \theta ~-~ \phi$ and  $ {\phi}_{L}~=~~ \theta ~+~ \phi$.
The total Hamiltonian is
\bea
H & = & {H_0}
+ \frac{4 E_{Z12}}{{(2 \pi \alpha)}^2} \int ~dx :cos(4 \sqrt{K} \phi (x)):
\nonumber\\
& & + \frac{ E_{C0}}{\pi \alpha} \int  ({{\partial}_x} \phi (x))~dx
\eea
$H_0$ 
is the non-interacting part of the Hamiltonian,
The dissipative strength (${\alpha}_{1} $) of this system in the absence of
Co-tunneling effect is
\beq
 {{\alpha}_1} ~  =~\frac{1}{4} \sqrt{ \frac{2 E_{J1} }
{2 {E_{J1}} ~+~\frac{16 E_{Z1} }{\pi} } }
\eeq
If we consider the total effect of Co-tunneling process, the system reduces
to Heisenberg spin chain with NN and NNN interactions. In this limit the
dissipative strength (${{\alpha}_2}$) of the system is
\beq
 {{\alpha}_2}  ~ =~\frac{1}{4} \sqrt{ \frac{ 2 E_{J1} }
{E_{J1}~+~\frac{4}{\pi} (4 E_{Z1}~ -~\frac{3 {E_{J1}}^2 }{4 E_{C0}})} }
\eeq
We calculate the dissipation strength by calculating K for both cases and
then we use the relation $K = 4 {\alpha}$.
This is the first analytical derivation of flux induced dissipation strength
in terms of the interactions of the system. We consider these two processes to
emphasis the importance of Co-tunneling effect for this system.

\noindent
{\bf Physical Analysis of Renormalization Group Equation and 
Calculation of Tunneling Resistance at The Quantum Critical Point:}\\
The RG equation based on the Eq. (7) is, 
\beq
\frac{d z}{d lnb}~=~ (1 - {\alpha}') z
\eeq
Following Ref.\cite{suj2}, we can write
fugacity 
depends
on length scale and temperature as,
$ z(L)~\propto L^{1 - {\alpha}'}$, ${z} (T) ~\propto {T}^{ {{\alpha}'} ~-~1 }$.
The physical explanations
based on this RG equation are in order.
\\
In our study, the resistance is evolving due to dissipation
effect at very low
temperature (few mili Kelvin, 
less than the superconducting Coulomb blocked temperature).
According to our calculations, for large dissipation ($ {\alpha}' > 1$),
$ R(T) \propto  R_Q {T^{{\beta}_1}}$ and ${{\beta}_1} > 0$.
Therefore at very low temperature, the system shows
stable SC behaviour (region B of Fig. 1). 
and the system shows no more saturated resistance behaviour
(region D of Fig. 1).
When ${\alpha}' < 1$,
the resistance of system
$ R(T) \propto  R_Q  {T^{{- \beta}_2}}$ and
${{\beta}_2} > 0 $. So at very low temperature,
the resistance of the system shows Kondo-like divergence behavior
(region A of Fig. 1).
These stable SC and Kondo behaviour
are unseen in the experimental findings because they have measured
the zero bias resistance at around $40-50$ mK.
According to our calculations, for large dissipation (${\alpha}' > 1$),
$ R(T) \propto R_Q {L^{- {\gamma}_1}}$ and ${{\gamma}_1} > 0$.
Therefore the longer array system shows
the less resistive state than shorter array in the
superconducting phase. When ${\alpha}' < 1$,
the resistance of system
$ R(T) \propto R_Q {L^{{ \gamma}_2}}$,
where ${{\gamma}_2} > 0 $
($ {{\beta}_1}, {{\beta}_2}, {{\gamma}_1}$ and ${{\gamma}_2}$
are independent numbers).
So the resistance in the insulating state is larger for longer
array system than shorter one.
We find the dual behavior of the resistance 
for lower and higher values of magnetic field.\\
From the knowledge of LL physics, we know that for a
particular range of $K$, there is a metallic state in between the insulating
phase and the superconducting phase of the system.
The range of $K$ depends on the nature of the system.
Therefore we conclude that
there is also a metallic state for a particular range of magnetic flux induced
dissipation strength.
This prediction is unnoticed in the experimental findings.
At
very low temperature around $\alpha \sim 1 $ 
(region E of Fig. 1), the quantum phase slip
centres poliferate and as a result a screening appears in the
system. A detailed explanation of QPS for this
type of system is discussed in Ref. \cite{suj2}. Physical explanation is
as follows.\\
Here we consider two QPS with co-ordinates (${x_1} , {\tau}_1 $) and
(${x_2} , {\tau}_2 $), we assume that the cores of the QPS centres donot
overlap, i.e., $| {x_2} - {x_1}| > {x_0} (=\xi) $ and
$| {\tau}_2 - {\tau}_1 | > {{\tau}_0} (= \frac{1}{\Delta}) $.
Where $\xi$ and $\Delta$ are the coherence length and SC orderparameter
of the system.
We have
already proven that the topological charges interact with each other
logarithmically. Therefore we can write the action of the system as
$ S_{QPS} = 2 S_{core} - {\mu} {{e}_1} {{e}_2} ln |{x_1} - {x_2}| $
. QPS with opposite topological charge attract each other and same
charge repel with each other. Suppose we consider a gas of $n$ QPS
and assume that QPS cores do not overlap. When a current is passing through
the system, we can write the total action of the system as
$ {S_{QPS}^n} = n S_{core} + {S_{int}} $ \cite{zai1}.
Where,
\beq
{S_{int}} = -{\mu} \sum_{i \neq j} {e_i}{e_j} ln( \frac{{\rho}_{ij}}{x_0})
+ \sum_{i} \frac{{\Phi}_{0} I}{c} {e_i}{{\tau}_i}
\eeq
Where 
${\rho}_{ij}$ is the distance between the two QPS at the site $i$ and $j$ 
and
$I$ is the current passing through the system. The grand partition function of
the system is represented as a sum of all topological charges and the
analytical form is equivalent to Eq.6, $z \sim e^{-S_{core}} $.
In the absence of current, Eq. (14) 
define the model for a 2-dimensional Coulomb gas interacting logarithmically.
We consider the very low temperature limit ($T \rightarrow 0$). 
The RG equations of this
2-dimensional Coulomb gas are the following
$\frac{\partial \mu}{\partial l} = - 4 {{\pi}^2} {{\mu}^2} z^2 $,
$\frac{\partial z}{\partial l} = (2 - {\mu} )z $. In our case we can write the
second equation as
$\frac{\partial z}{\partial log (\frac{\Delta}{T} ) } = (2 - {\mu})z $.
Finally it yields $ z (T) = z (\Delta) {( \frac{\Delta}{T})}^{2 -\mu}$.
When the interaction coefficient is less than 2, then the RG equation
diverges at the scale $ {T^*}$ (= $ {z}^{\frac{1}{2 -\mu}}{\Delta} $).
At the scale larger than $\xi$ the interaction between the QPS is screened
and cease to be logarithmic, it becomes an exponentially decaying function
($\sim e^{-2 {\mu} {K_0} (\sqrt{x^2 + {\tau}^2}) }$), ${K_0}$ is
the modified Bessel function.\\
\begin{figure}
\includegraphics[scale=0.50,angle=0]{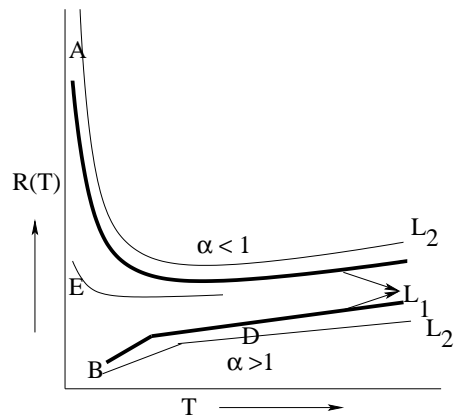}
\caption{ Schematic phase diagram showing the variation of resistance
with temperature based on the renormalization group study.
Region A is the Kondo like
behaviour of the system, region B is superconducting phase,
region D is the saturated 
finite resistance and region E is the metallic phase. ${L_2} > {L_1}$ }
\label{Fig. 2 }
\end{figure}
When ${\alpha}' = 1$, i.e.,
$ {\Phi} ~=({{\Phi}_0}/{\pi})cos^{-1} (\frac{R_s}{R_Q}) $,
the system has no length scale dependence SI 
transition at very low temperature. This is the critical
behavior of system for a specific value of magnetic field.
The analytical expression for tunneling resistance at the 
quantum critical point can be calculated by comparing the
expression of ${\alpha}_1 $ and ${\alpha}_2 $
with the expression of ${\alpha}^{'}$. 
${R_S^{(1)}}= R_{Q} ( \frac{16 E_{Z1}}{30 \pi E_{J1}})$ and
${R_S^{(2)}}= R_{Q} (\frac{5 \pi E_{C0}}{4 E_{J1}}) 
\sqrt{1 + \frac{2.77 E_{Z1} }{ E_{C0} }}$ are the tunneling
resistance at the quantum critical point in the absence and 
presence of Co-tunneling effect respectively. It is clear from our analytical
derivation that the tunneling resistance at quantum critical point
is not $R_Q $ but it is rather non-universal. $R_S^{(2)} $ is proportional
to the $E_{C0}$ as one would expect from the physical criteria of the
system.   
\\
Here we discuss the importance of Co-tunelling effect explicitly. 
In the SC phase
when ${\alpha }> 1 $, from the analysis of ${\alpha }_1 $, we get
the condition
$ \frac{- 16 E_{Z1}}{\pi} > 30  {E_ {J10} }
|cos (\frac{\pi \Phi}{{\Phi}_{0}})|  $. This condition is unphysical,
because the all coupling constants are repulsive. 
In presence of Co-tunneling
process, from the analysis ${\alpha}_2$, 
we achieve the condition of SC 
$ {E_{J10} } |cos (\frac{\pi \Phi}{{\Phi}_{0}})|
(\frac{192}{4 \pi {E_{C0}}} {E_{J10} } |cos (\frac{\pi \Phi}{{\Phi}_{0}})|
- 30 ) > \frac{256}{\pi} $. This condition is physically reliable
. Similarly one can do the 
analysis for the insulating phase for
both absence and presence of co-tunneling effect.\\
{\bf Conclusions:} we have found the non-universal tunneling
resistance at the quantum critical point of mesoscopic SQUIDs
array. We have found stable SC and Kondo phase at very low temperature,
which is still unseen experimentally. We have found 
the length scale independent metallic phase in
the system. The importance of Co-tunneling effect has been studied explicitly.\\
{\bf Acknowledgement:} The author would like to acknowledge CCMT of the
Physics Department of IISC for extended facility and Dr. S. Vidyadhiraja for
several important comments on this work and finally Dr. R. Srikanth 
and Dr. B. Murthy for reading
the manuscript very critically.

\end{document}